\documentclass[%
reprint,
superscriptaddress,
amsmath,amssymb,
aps,
prl,
floatfix
]{revtex4-2}

\usepackage{graphicx}
\usepackage{siunitx}  
\usepackage{booktabs}
\usepackage{hyperref}
\usepackage{tabularx}

\begin{document}
\title{A Sideways Crack-Path Instability Amplified by Slow Loading}

\author{G. Pallares}
\affiliation{CESI LINEACT, Montpellier campus, F-34130 Mauguio, France}
\affiliation{Universit\'e Lyon 1, CNRS, ILM, UMR 5306, Villeurbanne, France}
\author{O. Ramos}
\author{L. Vanel}
\email{Corresponding author: loic.vanel@univ-lyon1.fr}
 \affiliation{Universit\'e Lyon 1, CNRS, ILM, UMR 5306, Villeurbanne, France}

\date{today}

\begin{abstract}
Crack-path instabilities in solids are typically associated with high propagation velocities. In contrast, we show that nanoparticle-filled elastomers exhibit a sideways crack-path instability that is amplified as the pulling velocity decreases and nearly vanishes at high pulling velocities.
The sideways deviation amplitude decreases logarithmically with increasing pulling velocity and scales as the square root of the filler fraction. Large deviations develop at low mean crack velocities and are associated with effective fracture energies up to two orders of magnitude lower. These findings are consistent with subcritical rupture kinetics and suggest that the deviation amplitude is limited by the elasto-adhesive length.
\end{abstract}

\maketitle


In homogeneous isotropic solids under Mode-I loading, cracks are generally expected to propagate along a straight path perpendicular to the applied load. At high propagation velocities, this path can become dynamically unstable, leading to oscillatory crack trajectories or microbranching \cite{Bouchbinder2014,Chen2017,Kolvin2017}. In heterogeneous brittle materials, disorder can also produce intermittent crack propagation and rough fracture surfaces \cite{Santucci2004,Mallick2007,Bonamy2011}, which may have an impact on fracture energy \cite{Scheibert2010}. Cracks may also be deflected by structural anisotropy toward directions of lower fracture resistance or greater energy release rate \cite{Hakim2005}. In some tough elastomers, the initially isotropic structure can become anisotropic under large stretch, allowing cracks to propagate sideways, approximately parallel to the loading direction \cite{Gent2003, Lee2019}. However, the influence of loading rate on crack-path selection remains unclear, as contrasting rate dependences have been reported for sideways crack propagation in elastomers \cite{Gherib2010,Lee2019}.

In elastomers, crack-path deviations have been observed under various loading conditions. Under tearing conditions, both loading rate and temperature can strongly influence the stability of the crack trajectory \cite{Greensmith1956,Stacer1985a,Stacer1985b}, while biaxial loading and the resulting molecular orientation may affect sideways crack propagation \cite{Marano2010}. Depending on the material and loading conditions, deviations may remain small and produce rough fracture surfaces \cite{Tsunoda2000,Gabrielle2011}, or develop into large departures from the conventional crack path \cite{Lake1991,Lee2019}. Three-dimensional effects may further complicate this behavior in thick elastomer samples \cite{Lee2019,Xue2024}. Despite these observations, systematic quantitative measurements of the amplitude and growth dynamics of sideways cracks remain scarce \cite{Gabrielle2011}.

Here, we systematically quantify sideways crack propagation in nanoparticle-filled elastomers over broad ranges of pulling velocity and filler fraction. We find that, in marked contrast to conventional high-speed crack-path instabilities, this sideways crack-path instability is amplified as the loading rate decreases, while it nearly vanishes at high loading velocities. The deviation amplitude decreases logarithmically with increasing pulling velocity, while the prefactor of this dependence increases approximately as the square root of the filler fraction. Measurements of the growth time show that large deviations develop at much lower mean crack velocities and are associated with effective fracture energies reduced by up to two orders of magnitude. We thus identify a time-dependent crack-path instability whose direction may be selected by stretch-induced structural anisotropy and whose amplitude is controlled by subcritical rupture kinetics. Remarkably, the deviation amplitude always remains below the elasto-adhesive length.

\begin{figure}[!htbp]
\centering
\includegraphics[width=8.2cm]{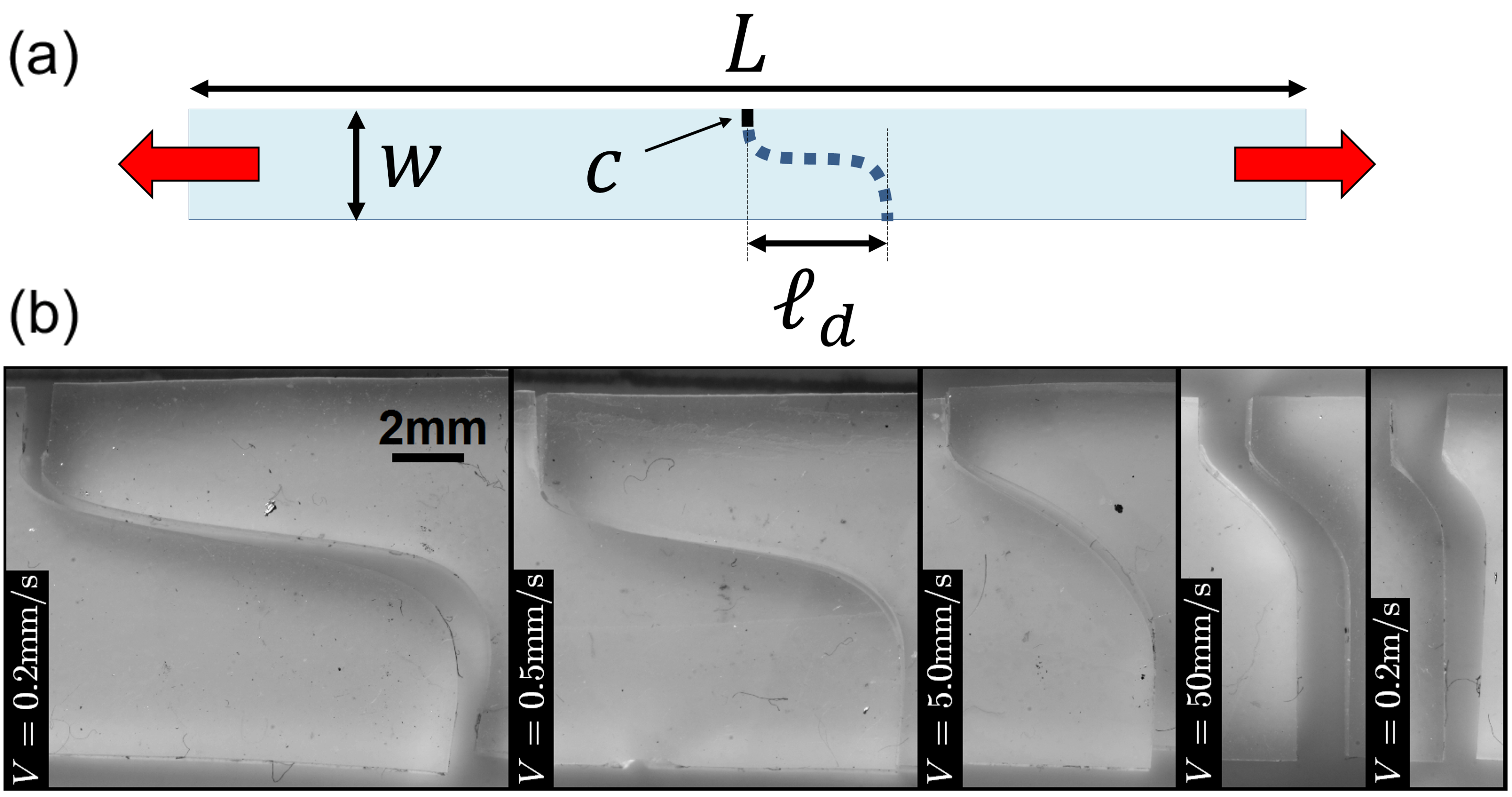}
\caption{(a) Schematic drawing of the test used to measure crack deviations $\ell_d$. The arrows indicate the loading direction. (b) Typical post-mortem images illustrating these deviations in the notched Si23 sample fractured at pulling velocities from $V = 0.2$~mm.s$^{-1}$ to $V = 200$~mm.s$^{-1}$ (scale bar: $2$~mm).}
\label{fig:CrackObs}
\end{figure}

Single edge notch tensile (SENT) specimens with dimensions of length $L = 100$~mm, width $w = 10$~mm and thickness $t=2$~mm  were prepared with a pre-crack of length $c = 1.6 \pm 0.1$~mm to control crack initiation (Fig.~\ref{fig:CrackObs}). We investigated five PDMS-based elastomer formulations with different silica mass fractions $C$, prepared specifically for this study by Elkem Silicones (Saint-Fons, France)  (Table~\ref{tab:Formulations}).
\begin{table}[ht]
\centering
\begin{tabular}{ c c c c c c}
  \hline
 Formulation &  Si0 & Si5 & Si9 & Si17 & Si23\\
  \hline
Rhodorsil Gum 759 (w\%) & 100 & 95 & 91 & 83 & 77\\
Silica Ae55 (w\%)      &   0 &  5 &  9 & 17 & 23\\
  \hline
\end{tabular}
\caption{Sample formulations, given as the mass fraction of silicone (Rhodorsil\textsuperscript{\textregistered} Gum 759) and fumed silica nanoparticles (AE55, Elkem Silicones).}
\label{tab:Formulations}
\end{table}
The samples were stretched using a custom-built tensile apparatus at constant velocities spanning three orders of magnitude, from $V = 0.2$~mm.s$^{-1}$ to $V = 200$~mm.s$^{-1}$, corresponding to tensile strain rates between $\dot{\epsilon} = 2 \times 10^{-3}$~s$^{-1}$ and $2 $~s$^{-1}$ . Once clamped between the machine jaws, each specimen was subjected to a low pre-stress of $0.05$~MPa. The applied force was recorded at $100$~Hz with an accuracy of $\pm 0.01$~N. All tests were conducted at room temperature $T = 25^{o}\mathrm{C}$ until sample failure. Post-mortem, the sideways crack was imaged using a Canon EOS~100D camera equipped with a Canon EF-S $18-55$~mm lens, yielding a pixel resolution of $5.7$~$\mu$m. Representative fracture surface images for the Si23 formulation stretched at various pulling velocities show that crack deviations from a straight path are much larger at low velocities (Fig.~\ref{fig:CrackObs}).

\begin{figure}[!htbp]
\centering
\includegraphics[width=7cm]{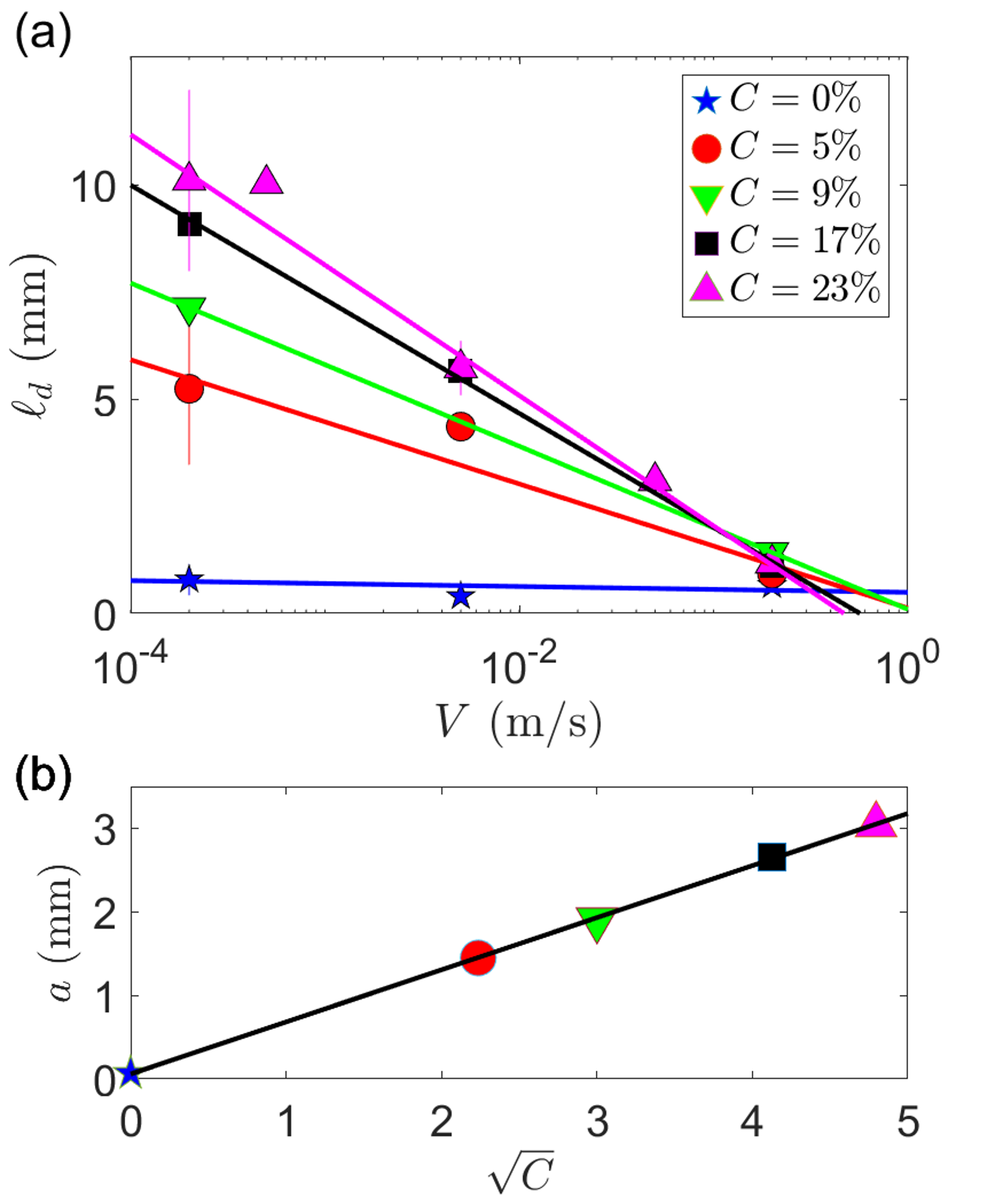}
\caption{(a) Deviation amplitude $\ell_d$ as a function of the loading velocity $V$. Lines are linear fits $\ell_d = a \ln(V/V_m)$. (b) Evolution of the fitted parameter $a$ as a function of the square root of filler mass fraction $C$.}
\label{fig:LxV}
\end{figure}

To quantify crack deviations, we measured the perpendicular distance $\ell_d$ between the initial notch and the final crack path, as defined in Fig.~\ref{fig:CrackObs}. Fig.~\ref{fig:LxV}(a) shows that, at a fixed pulling velocity, this deviation increases monotonically with filler mass fraction, in contrast to the nonmonotonic behavior observed for the tear energy \cite{Castro2015,SupplementalMaterial}. Conversely, for a given filler content, the crack deviation decreases markedly as the pulling velocity increases over three orders of magnitude. At sufficiently high velocities, deviations become negligible for all filler fractions, although a small residual deviation persists, as observed even in pure PDMS. The observed behavior is approximately described by:
\begin{equation}
\ell_d \simeq a \ln(V_m/V),
\label{eq:elldV}
\end{equation}
where $V_m$ is a characteristic velocity above which deviations almost vanish. Strikingly, the prefactor $a$ increases with the square-root of the filler fraction:  $a \simeq 0.62\sqrt{C}$ (Fig.~\ref{fig:LxV}(b)).

\begin{figure}[!htbp]
\centering
\includegraphics[width=7cm]{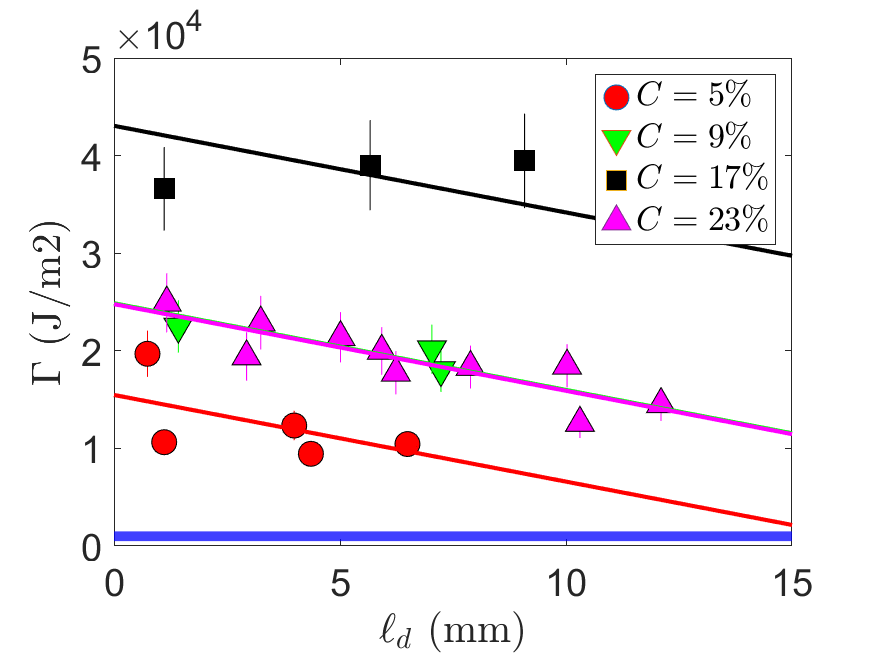}
\caption{Tear energy $\Gamma$ of pre-notched samples as a function of deviation amplitude $\ell_d$. Lines are guides to the eye, drawn with the slope obtained from the fit at $C=23\%$. The blue horizontal line indicates the tear-energy level of pure PDMS.}
\label{fig:GammaeLx}
\end{figure}

When comparing the measured fracture energy $\Gamma$ \cite{SupplementalMaterial} with the crack deviation amplitude $\ell_d$ (Fig.~\ref{fig:GammaeLx}), we observe a general tendency of the tear energy to decrease with increasing deviation. This trend is especially obvious at the highest filler fraction ($C=23\%$). These results challenge the common assumption that crack deviation enhances material strength; on the contrary, they suggest that deviations correlate with reduced tear resistance.

\begin{figure}[!htbp]
\centering
\includegraphics[width=8cm]{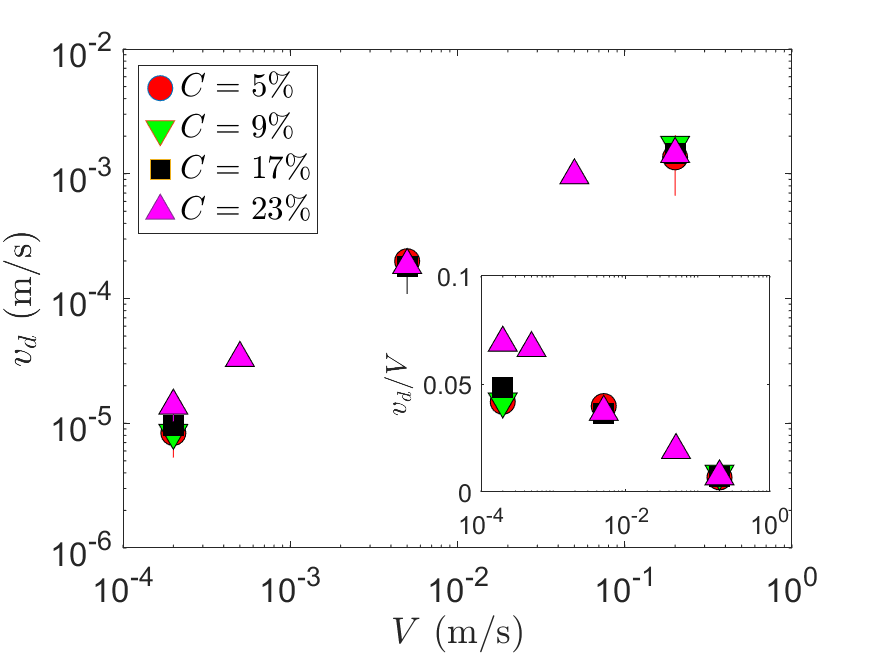}
\caption{Sideway crack velocity $v_d$ as a function of the imposed displacement velocity ranging from $V = 0.2$~mm.s$^{-1}$ to $V = 200$~mm.s$^{-1}$. Inset: Normalized sideways crack velocity.}
\label{fig:vcV}
\end{figure}

To better understand how crack deviations develop and why they tend to reduce the overall tear energy, we first estimate their mean growth velocity as $v_d = \ell_d / \tau$, where $\tau$ is the time to rupture. We find that the deviation velocity is one to two orders of magnitude lower than the imposed pulling velocity (Fig.~\ref{fig:vcV}). As shown in the inset of Fig.~\ref{fig:vcV}, the ratio $v_d/V$ is larger at lower pulling velocities, indicating that crack deviations grow more readily under slow loading. This suggests that time-dependent rupture processes occurring under subcritical conditions, below the nominal failure threshold, contribute to the growth of the deviations. Such processes are typically thermally activated and may enable fracture at lower tear energies.

\begin{figure}[!htbp]
\centering
\includegraphics[width=7cm]{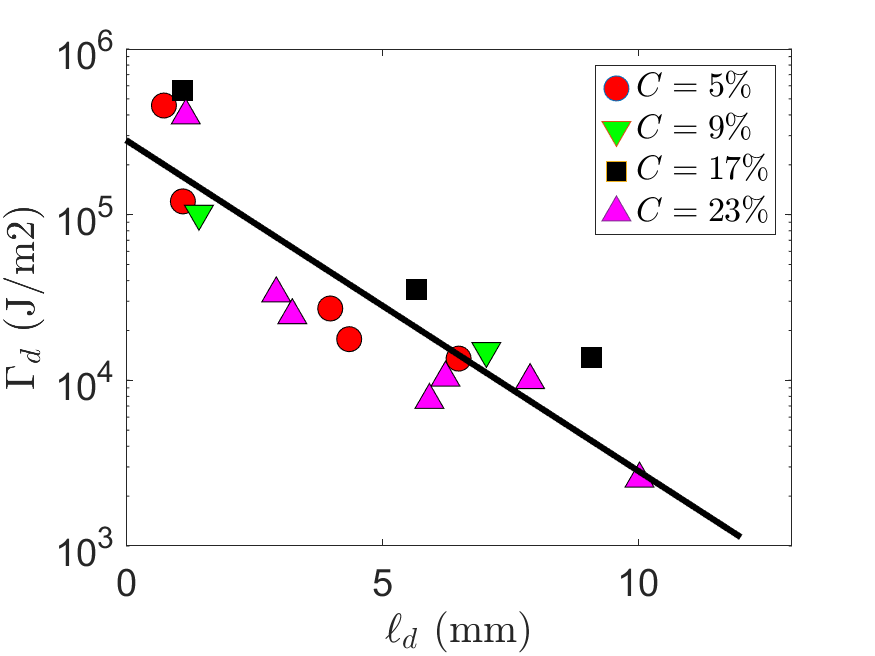}
\caption{Fracture energy $\Gamma_d$ associated to sideways crack growth as a function of the crack deviation amplitude $\ell_d$.}
\label{fig:Gammadelld}
\end{figure}

To estimate the fracture energy $\Gamma_d$ specifically associated with sideways crack growth, we determine the mechanical energy released during this process \cite{SupplementalMaterial}. We find that $\Gamma_d$ decreases with increasing deviation amplitude, with long deviations exhibiting values one to two orders of magnitude lower than short deviations (Fig.~\ref{fig:Gammadelld}). This result is consistent with large deviations developing through time-dependent rupture mechanisms that become accessible when sufficient time is available. For long deviations, the fracture energy approaches that of the unfilled silicone \cite{SupplementalMaterial}, thereby effectively suppressing the reinforcing effect of the filler. Although $\Gamma_d$ strongly depends on the deviation amplitude, it exhibits no clear dependence on filler content. Its variation is approximately described by \begin{equation} \Gamma_d=\Gamma_m \exp\left(-\ell_d/\delta\right), \label{eq:Gammadelld} \end{equation} with $\delta = 2.17\,$mm and $\Gamma_m \simeq 2.8\times 10^{5}\,$J\,m$^{-2}$.

Combining Eq.\~ (\ref{eq:elldV}) and (\ref{eq:Gammadelld}) leads to a power-law dependence on pulling velocity:
\begin{equation}
\Gamma_d=\Gamma_m \left(\frac{V}{V_m}\right)^\alpha,
\end{equation}
where $\alpha=a/\delta \approx 0.29-0.61$. Such small exponents are typical of polymer fracture dynamics and are commonly attributed to rate-dependent rupture processes \cite{Greensmith1956,Lake1991,Maugis1978,Stacer1985c,Gent1996,Barquins1997a}. The exponent $\alpha$ has been reported to increase with filler content~\cite{Morishita2016}, consistently with the increase in the parameter $a$ with filler fraction $C$ observed here. Thus, the power-law dependence of the fracture energy $\Gamma_d$ on the pulling velocity $V$ further supports the interpretation of crack deviations as a manifestation of a rate-dependent growth process. Furthermore, recent simulations of hyperelastic materials show that the elastic energy release rate $G$ for sideways cracks decreases with increasing crack length~\cite{Lee2019}. A decrease in $\Gamma_d$ with deviation amplitude, as observed experimentally (Fig.~\ref{fig:Gammadelld}), is therefore necessary to allow the propagation of large sideways cracks despite the decreasing energy release rate.

A central quantity governing fracture in soft polymers is the elasto-adhesive length, defined as $\ell_{\mathrm{EA}}=\Gamma/E$, where $E$ is the elastic modulus of the elastomer \cite{Creton2016}. This length compares the energy required to create new surfaces with the elastic energy density and sets the spatial extent of the region surrounding the crack tip in which elastic strains become of order unity. Consequently, even in simple elastomers such as unfilled PDMS, linear elasticity necessarily breaks down at distances comparable to $\ell_{\mathrm{EA}}$.

\begin{figure}[!htbp]
\centering
\includegraphics[width=8cm]{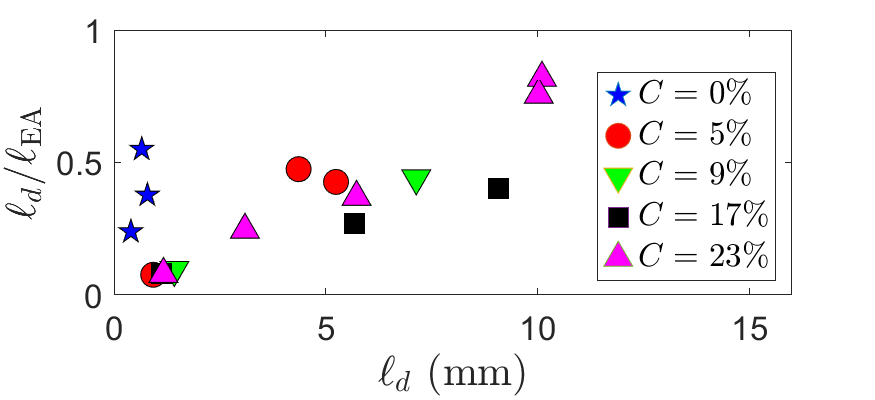}
\caption{Crack deviation amplitude $\ell_d$ normalized by the elasto-adhesive length $\ell_{\mathrm{EA}}$ as a function of $\ell_d$.}
\label{fig:ldlea}
\end{figure}

For each filled formulation, the crack-path deviation amplitude $\ell_d$ is systematically smaller than the elasto-adhesive length $\ell_{\mathrm{EA}}^{\mathrm{filled}}$ calculated using the fracture energy of the corresponding filled elastomer (Fig.~\ref{fig:ldlea} and \cite{SupplementalMaterial}). However, the two quantities are not directly correlated. While $\ell_{\mathrm{EA}}^{\mathrm{filled}}$ depends only weakly on the pulling velocity, $\ell_d$ decreases markedly with increasing pulling velocity. Their dependences on filler content also differ. At the highest filler mass fraction, $C=23\%$, $\ell_{\mathrm{EA}}^{\mathrm{filled}}$ decreases, whereas $\ell_d$ continues to increase. Thus, $\ell_{\mathrm{EA}}^{\mathrm{filled}}$ does not determine the deviation amplitude, although the systematic inequality $\ell_d<\ell_{\mathrm{EA}}^{\mathrm{filled}}$ suggests that it may provide an upper bound.

A distinct set of observations concerns pure PDMS. Small crack deviations of approximately $1~\mathrm{mm}$ are observed even without fillers (Fig.~\ref{fig:LxV}), consistent with their amplitude being bounded by the elasto-adhesive length of pure PDMS, $\ell_{\mathrm{EA}}^{\mathrm{pure}}\simeq1.5\pm0.5~\mathrm{mm}$ (Fig.~\ref{fig:ldlea}). Furthermore, the characteristic length $\delta$ introduced by Eq.~(\ref{eq:Gammadelld}) is approximately independent of filler content and is comparable to $\ell_{\mathrm{EA}}^{\mathrm{pure}}$. The effective fracture energy associated with large sideways cracks also approaches that of pure PDMS \cite{SupplementalMaterial}. These observations suggest that, over sufficiently long timescales, sideways crack growth in the filled elastomers becomes increasingly influenced by the fracture response of the pure PDMS matrix. The physical origin of the similarities between these characteristic length scales nevertheless remains unclear.

Although subcritical rupture kinetics appear to control the amplitude of the sideways crack instability, they do not explain why cracks propagate along the loading direction rather than perpendicular to it. Several mechanisms may account for this directional selection. First, recent studies have shown that solid inclusions modify the local strain-energy density, favoring propagation along directions of maximum energy density, which may differ from the expected Mode I direction~\cite{RozenLevy2020}. Second, experiments on strain-crystallizing elastomers have shown that polymer dynamics accelerate in regions ahead of sideways cracks, promoting growth along specific directions aligned with the loading direction~\cite{Ju2023}.

At a fixed pulling velocity, the rupture time does not vary much with filler fraction because the macroscopic stretch at break is similar for the different formulations. Thus, the increase in deviation amplitude with filler fraction cannot be attributed primarily to a longer time available for sideways crack growth. However, higher filler fraction will concentrate more deformation in the polymer matrix, causing some regions between filler aggregates to reach larger local stretches. This could allow subcritical sideways cracks to propagate farther within approximately the same rupture time. This interpretation remains qualitative and does not explain the observed square-root dependence on filler fraction.

In conclusion, our results show that sideways crack deviations in nanoparticle-filled elastomers are strongly influenced by both filler content and pulling velocity. Increasing the filler fraction produces larger deviation amplitudes, whereas increasing the pulling velocity suppresses the instability, with deviations nearly vanishing at high velocities. Contrary to the common expectation that crack deflection enhances toughness, the macroscopic tear energy decreases as the deviation amplitude increases. The fracture energy associated with sideways propagation also decreases with the mean crack velocity and approaches that of pure PDMS for the slowest-growing cracks. This result may have important implications for the long-term durability of filled elastomers since their resistance to slow rupture may ultimately be governed by the fracture properties of the polymer matrix rather than those of the reinforced material.

\begin{acknowledgments}
The authors acknowledge the support of the French National Research Agency (ANR) under grant ANR-12-BS09-014 (StickSlip), and thank Elkem Silicones for providing filled elastomer samples. All data supporting the findings of this study are openly available here~\cite{Dataset}.
\end{acknowledgments}

\vspace{-5mm}

\bibliographystyle{apsrev4-2}
\bibliography{article}

@misc{SupplementalMaterial,
  note = {See Supplemental Material at [URL] for details of the experimental characterizations.}
}

@dataset{Dataset,
  author       = {Pallares, G. and Ramos, O. and Vanel, L.},
  title        = {Dataset - Aging in the Flow Dynamics of Dense Suspensions of Contactless Microparticles},
  month        =  oct,
  year         = 2026,
  publisher    = {Zenodo},
  doi          = {10.5281/zenodo.23361968 },
  url          = {https://doi.org/10.5281/zenodo.23361968 },
}

@Article{Creton2016,
  author    = {Creton, C. and Ciccotti, M.},
  journal   = {Rep. Prog. Phys.},
  title     = {Fracture and adhesion of soft materials: a review},
  year      = {2016},
  month     = apr,
  number    = {4},
  pages     = {046601},
  volume    = {79},
  af        = {Creton, CostantinoEOLEOLCiccotti, Matteo},
  doi       = {10.1088/0034-4885/79/4/046601},
  ei        = {1361-6633},
  pm        = {27007412},
  ri        = {CRETON, COSTANTINO/B-4077-2008},
  sn        = {0034-4885},
  tc        = {3},
}

@Article{Bouchbinder2014,
  author    = {Bouchbinder, Eran and Goldman, Tamar and Fineberg, Jay},
  journal   = {Reports on Progress in Physics},
  title     = {The dynamics of rapid fracture: instabilities, nonlinearities and length scales},
  year      = {2014},
  issn      = {1361-6633},
  month     = Mar,
  number    = {4},
  pages     = {046501},
  volume    = {77},
  doi       = {10.1088/0034-4885/77/4/046501},
  publisher = {IOP Publishing},
}

@Article{Chen2017,
  author   = {Chen, Chih-Hung and Bouchbinder, Eran and Karma, Alain},
  journal  = {Nature Physics},
  title    = {Instability in dynamic fracture and the failure of the classical theory of cracks},
  year     = {2017},
  issn     = {1745-2481},
  number   = {12},
  pages    = {1186--1190},
  volume   = {13},
  doi      = {10.1038/nphys4237},
  url      = {https://doi.org/10.1038/nphys4237},
}

@Article{Kolvin2017,
  author    = {Kolvin, Itamar and Fineberg, Jay and Adda-Bedia, Mokhtar},
  journal   = {Physical Review Letters},
  title     = {Nonlinear Focusing in Dynamic Crack Fronts and the Microbranching Transition},
  year      = {2017},
  issn      = {1079-7114},
  month     = Nov,
  number    = {21},
  pages     = {215505},
  volume    = {119},
  doi       = {10.1103/PhysRevLett.119.215505},
  publisher = {American Physical Society (APS)},
}

@Article{Bonamy2011,
  author    = {Bonamy, D. and Bouchaud, E.},
  journal   = {Physics Reports},
  title     = {Failure of heterogeneous materials: A dynamic phase transition?},
  year      = {2011},
  issn      = {0370-1573},
  month     = Jan,
  number    = {1},
  pages     = {1--44},
  volume    = {498},
  doi       = {10.1016/j.physrep.2010.07.006},
  publisher = {Elsevier BV},
}

@Article{Mallick2007,
  author    = {Mallick, N. and Cortet, P. P. and Santucci, S. and Roux, S. G. and Vanel, L.},
  journal   = {Physical Review Letters},
  title     = {Discrepancy between subcritical and fast rupture roughness: A cumulant analysis},
  year      = {2007},
  issn      = {1079-7114},
  month     = jun,
  number    = {25},
  pages     = {255502},
  volume    = {98},
  af        = {Mallick, N.EOLEOLCortet, P.-P.EOLEOLSantucci, S.EOLEOLRoux, S. G.EOLEOLVanel, L.},
  doi       = {10.1103/PhysRevLett.98.255502},
  publisher = {American Physical Society (APS)},
}

@Article{Santucci2004,
  author    = {Santucci, S. and Vanel, L. and Ciliberto, S.},
  journal   = {Physical Review Letters},
  title     = {Subcritical statistics in rupture of fibrous materials: Experiments and model},
  year      = {2004},
  issn      = {1079-7114},
  month     = aug,
  number    = {9},
  pages     = {095505},
  volume    = {93},
  doi       = {10.1103/PhysRevLett.93.095505},
  publisher = {American Physical Society (APS)},
}

@Article{Scheibert2010,
  author    = {Scheibert, J. and Guerra, C. and Célarié, F. and Dalmas, D. and Bonamy, D.},
  journal   = {Physical Review Letters},
  title     = {Brittle-Quasibrittle Transition in Dynamic Fracture: An Energetic Signature},
  year      = {2010},
  issn      = {1079-7114},
  month     = Jan,
  number    = {4},
  pages     = {045501},
  volume    = {104},
  doi       = {10.1103/PhysRevLett.104.045501},
  publisher = {American Physical Society (APS)},
}

@Article{Hakim2005,
  author    = {Hakim, Vincent and Karma, Alain},
  journal   = {Phys. Rev. Lett.},
  title     = {Crack Path Prediction in Anisotropic Brittle Materials},
  year      = {2005},
  issn      = {1079-7114},
  month     = {Dec},
  number    = {23},
  pages     = {235501},
  volume    = {95},
  doi       = {10.1103/PhysRevLett.95.235501},
  issue     = {23},
  publisher = {American Physical Society},
}

@Article{Stacer1985a,
  title                    = {Observations on the tearing of elastomers},
  author                   = {Stacer, R. G. and Yanyo, L. C. and Kelley, F. N.},
  journal                  = {Rubber Chem. Technol.},
  year                     = {1985},
  pages                    = {421--435},
  volume                   = {58},
}

@Article{STACER1985b,
  author    = {Stacer, R. G. and Kelley, F. N.},
  journal   = {Rubber Chem. Technol.},
  title     = {Criteria For Unstable Tearing Of Elastomers},
  year      = {1985},
  number    = {5},
  pages     = {924--938},
  volume    = {58},
}

@Article{Gent2003,
  author   = {Gent, A. N. and Razzaghi-Kashani, M. and Hamed, G. R.},
  journal  = {Rubber Chem. Technol.},
  pages    = {122-131},
  title    = {Why Do Cracks Turn Sideways?},
  volume   = {76},
  year     = {2003},
  doi      = {10.5254/1.3547727},
  issn     = {0035-9475},
  number   = {1},
  url      = {https://doi.org/10.5254/1.3547727},
}

@Article{Lee2019,
  author   = {Seunghyun Lee and Matt Pharr},
  journal  = {Proc. Natl. Acad. Sci.},
  pages    = {9251-9256},
  title    = {Sideways and stable crack propagation in a silicone elastomer},
  volume   = {116},
  year     = {2019},
  doi      = {10.1073/pnas.1820424116},
  number   = {19},
  url      = {https://www.pnas.org/doi/abs/10.1073/pnas.1820424116},
}

@Article{Gabrielle2011,
  Title                    = {Effect of Tear Rotation on Ultimate Strength in Reinforced Natural Rubber},
  Author                   = {Gabrielle, B. and Guy, L. and Albouy, P. A. and Vanel, L. and Long, D. R. and Sotta, P.},
  Journal                  = {Macromolecules},
  Year                     = {2011},
  Month                    = sep,
  Number                   = {17},
  Pages                    = {7006--7015},
  Volume                   = {44},
  Doi                      = {10.1021/ma2010926},
}

@Article{Tsunoda2000,
  author    = {Tsunoda, K. and Busfield, J. J. C. and Davies, C. K. L. and Thomas, A. G.},
  journal   = {J. Mater. Sci.},
  title     = {Effect of materials variables on the tear behaviour of a non-crystallising elastomer},
  year      = {2000},
  month     = oct,
  number    = {20},
  pages     = {5187--5198},
  volume    = {35},
  doi       = {10.1023/A:1004860522186},
}

@Article{Lake1991,
  author    = {Lake, G.J. and Samsuri, A. and Teo, S.C. and Vaja, J.},
  journal   = {Polymer},
  title     = {Time-dependent fracture in vulcanized elastomers},
  year      = {1991},
  number    = {16},
  pages     = {2963--2975},
  volume    = {32},
  doi       = {10.1016/0032-3861(91)90194-n}
}

@Article{Gherib2010,
  author   = {Gherib, S. and Chazeau, L. and Pelletier, J. M. and Satha, H.},
  journal  = {J. Appl. Polym. Sci.},
  pages    = {435-445},
  title    = {Influence of the filler type on the rupture behavior of filled elastomers},
  volume   = {118},
  year     = {2010},
  doi      = {https://doi.org/10.1002/app.31606},
  number   = {1},
  url      = {https://onlinelibrary.wiley.com/doi/abs/10.1002/app.31606},
}

@Article{Ju2023,
  author    = {Ju, Jianzhu and Sanoja, Gabriel E. and Nagazi, Med Yassine and Cipelletti, Luca and Liu, Zezhou and Hui, Chung Yuen and Ciccotti, Matteo and Narita, Tetsuharu and Creton, Costantino},
  journal   = {Phys. Rev. X},
  title     = {Real-Time Early Detection of Crack Propagation Precursors in Delayed Fracture of Soft Elastomers},
  year      = {2023},
  month     = may,
  pages     = {021030},
  volume    = {13},
  cites     = {Ju2023-1,Ju2023-2},
  doi       = {10.1103/PhysRevX.13.021030},
  issue     = {2},
  numpages  = {17},
  publisher = {American Physical Society},
  url       = {https://link.aps.org/doi/10.1103/PhysRevX.13.021030},
}

@Article{Maugis1978,
  author    = {Maugis, D. and Barquins, M.},
  journal   = {J. Phys. D-Appl. Phys.},
  title     = {Fracture Mechanics And Adherence Of Viscoelastic Bodies},
  year      = {1978},
  number    = {14},
  pages     = {1989--\&},
  volume    = {11},
}

@Article{Stacer1985c,
  Title                    = {Time-dependent tearing of carbon black filled and strain crystallizing vulcanizates},
  Author                   = {Stacer, R.G. and von Meerwall, E.D. and Kelley, F.N.},
  Journal                  = {Rubber Chem. Technol.},
  Year                     = {1985},
  Pages                    = {913--923},
  Volume                   = {58},
}

@Article{Gent1996,
  author    = {Gent, A. N.},
  journal   = {Langmuir},
  title     = {Adhesion and strength of viscoelastic solids. Is there a relationship between adhesion and bulk properties?},
  year      = {1996},
  month     = sep,
  number    = {19},
  pages     = {4492--4496},
  volume    = {12},
}

@Article{Barquins1997a,
  author  = {Barquins, M. and Ciccotti, M.},
  journal = {Int. J. Adhes. Adhes.},
  title   = {On the kinetics of peeling of an adhesive tape under a constant imposed load},
  year    = {1997},
  pages   = {65},
  volume  = {17},
}

@Article{Xue2024,
  author    = {Xue, Nan and Long, Rong and Dufresne, Eric R. and Style, Robert W.},
  journal   = {Phys. Rev. X},
  pages     = {011054},
  title     = {Elastomers {Fail} from the {Edge}},
  volume    = {14},
  year      = {2024},
  doi       = {10.1103/PhysRevX.14.011054},
  number    = {1},
  url       = {https://link.aps.org/doi/10.1103/PhysRevX.14.011054},
  urldate   = {2025-03-13},
  month     = mar,
  publisher = {American Physical Society},
}

@Article{Marano2010,
  author   = {Marano, Claudia and Calabrò, Roberto and Rink, Marta},
  journal  = {J. Polym. Sci., Part B: Polym. Phys.},
  pages    = {1509-1515},
  title    = {Effect of molecular orientation on the fracture behavior of carbon black-filled natural rubber compounds},
  volume   = {48},
  year     = {2010},
  doi      = {https://doi.org/10.1002/polb.22054},
  number   = {13},
  url      = {https://onlinelibrary.wiley.com/doi/abs/10.1002/polb.22054},
}

@Article{Castro2015,
  author   = {de Castro, J. G. and Zargar, R. and Habibi, M. and Varol, S. H. and Parekh, S. H. and Hosseinkhani, B. and Adda-Bedia, M. and Bonn, D.},
  journal  = {Appl. Phys. Lett.},
  title    = {Nonmonotonic fracture behavior of polymer nanocomposites},
  year     = {2015},
  issn     = {0003-6951},
  month    = {06},
  number   = {22},
  pages    = {221904},
  volume   = {106},
  doi      = {10.1063/1.4922287},
  url      = {https://doi.org/10.1063/1.4922287},
}

@Article{Morishita2016,
  author    = {Morishita, Yoshihiro and Tsunoda, Katsuhiko and Urayama, Kenji},
  journal   = {Phys. Rev. E},
  title     = {Velocity transition in the crack growth dynamics of filled elastomers: Contributions of nonlinear viscoelasticity},
  year      = {2016},
  month     = apr,
  pages     = {043001},
  volume    = {93},
}

@Article{RozenLevy2020,
  author    = {Rozen-Levy, Lital and Kolinski, John M. and Cohen, Gil and Fineberg, Jay},
  journal   = {Phys. Rev. Lett.},
  pages     = {175501},
  title     = {How Fast Cracks in Brittle Solids Choose Their Path},
  volume    = {125},
  year      = {2020},
  doi       = {10.1103/PhysRevLett.125.175501},
  issue     = {17},
  url       = {https://link.aps.org/doi/10.1103/PhysRevLett.125.175501},
  month     = oct,
  numpages  = {5},
  publisher = {American Physical Society},
}

@Article{Greensmith1956,
  author    = {Greensmith, H. W.},
  journal   = {J. Polym. Sci.},
  title    = {Rupture of rubber. IV. Tear properties of vulcanizates containing carbon black},
  year      = {1956},
  issn      = {0035-9475},
  month     = mar,
  number    = {98},
  pages     = {175--187},
  volume    = {21},
  doi      = {https://doi.org/10.1002/pol.1956.120219802},
  url      = {https://onlinelibrary.wiley.com/doi/abs/10.1002/pol.1956.120219802},
}

\newpage

\appendix

\onecolumngrid

\section{Supplemental Material}
For the various material formulations and test conditions, this Supplemental Material reports the stress and stretch ratio at failure measured on unnotched samples, the procedure used to estimate the fracture energy associated with sideways crack growth, and the calculated elasto-adhesive lengths.

\subsection{Fracture energy dependence on filler fraction and loading velocity}
\begin{figure}[!htbp]
\includegraphics[width=0.45\linewidth]{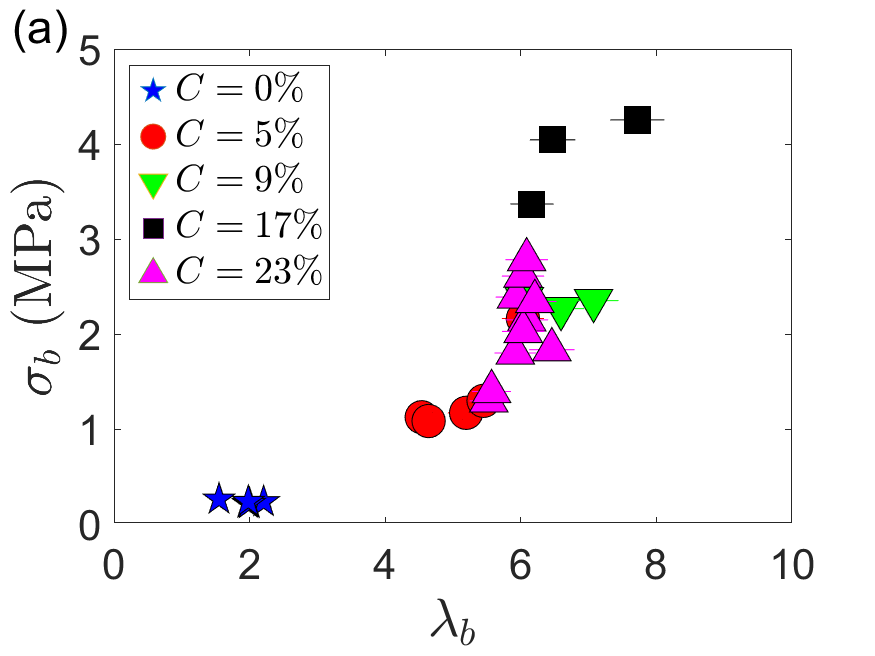}
\includegraphics[width=0.45\linewidth]{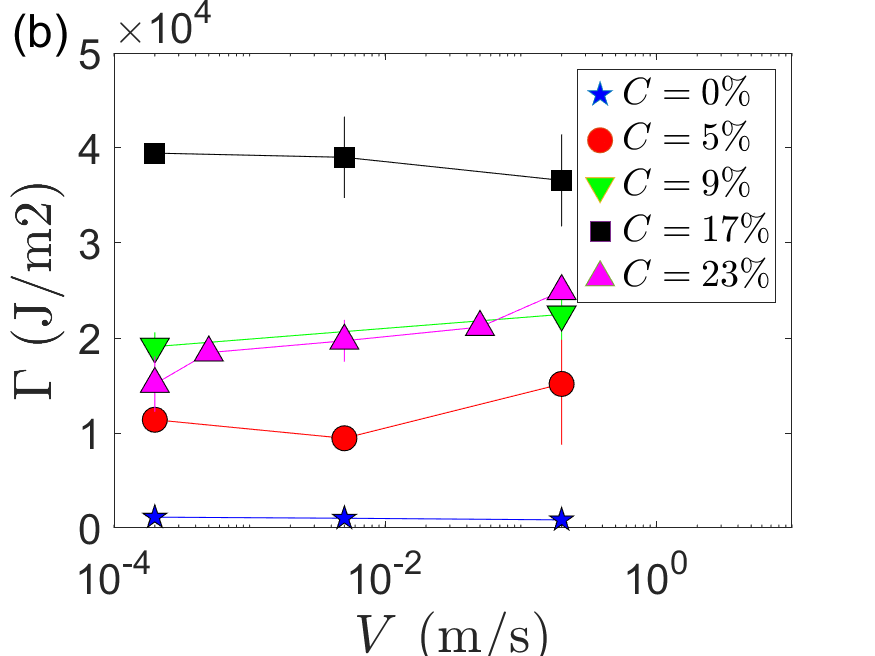}
\caption{(a) Stress-elongation ratio curves at the breaking point, $\sigma_b(\lambda_b)$, for pre-notched samples ($c$ = 1.6 $\pm$ 0.1mm) at different strain rates. (b) Tear energy, $\Gamma$, of the pre-notched samples, for different formulations and loading velocities ranging from $V = 0.2$~mm.s$^{-1}$ to $V = 200$~mm.s$^{-1}$}.
\label{fig:MasterBreakCurve}
\end{figure}

Stress versus stretch ratio data at the breaking point, $\sigma_b(\lambda_b)$,  are shown in Fig.~\ref{fig:MasterBreakCurve}a for pre-notched specimens. These data correspond to unfilled samples ($C=0\%$) and filled samples with silica mass fractions $C=\left\{5,9,17,23\right\} \%$, tested at room temperature over strain rates ranging from $\dot{\epsilon} = 2 \times 10^{-3}$~s$^{-1}$ to $2 $~s$^{-1}$. 
The addition of silica generally enhances the mechanical strength, although this effect becomes non-monotonic at high filler contents. In particular, at the highest filler content ($C=23\%$), the strength decreases to values comparable to those of moderately filled samples ($C=9\%$). This non-monotonic trend has also been reported in previous studies of filled elastomers \cite{Castro2015}.

To estimate the fracture energy $\Gamma$, we consider the energy release rate during crack advance, which is defined as the variation in mechanical energy per unit increase in fracture surface area. For a SENT specimen, $\Gamma$ is given by \cite{Lake1991}:
\begin{equation}
\Gamma = \frac{2\pi W_b c}{\sqrt{\lambda_b}},
\label{eq:EnergyReleaseRate}
\end{equation}
where $c$ is the initial notch length, $\lambda_b$ the stretch ratio in the uniformly deformed region and $W_b$ the elastic energy density in the same region (\textit{i.e.} far from the crack tip), both evaluated at the breaking point.
The values of $\Gamma$ as a function of tensile velocity $V$ for each formulation are shown in Fig.~\ref{fig:MasterBreakCurve}b. Unfilled samples have a typical tear energy of $10^3$J.m$^{-2}$, while filled samples exhibit tear energies 10 to 40 times larger.

\subsection{Fracture energy of crack deviations}

\begin{figure}[!htbp]
\centering
\includegraphics[width=0.45\linewidth]{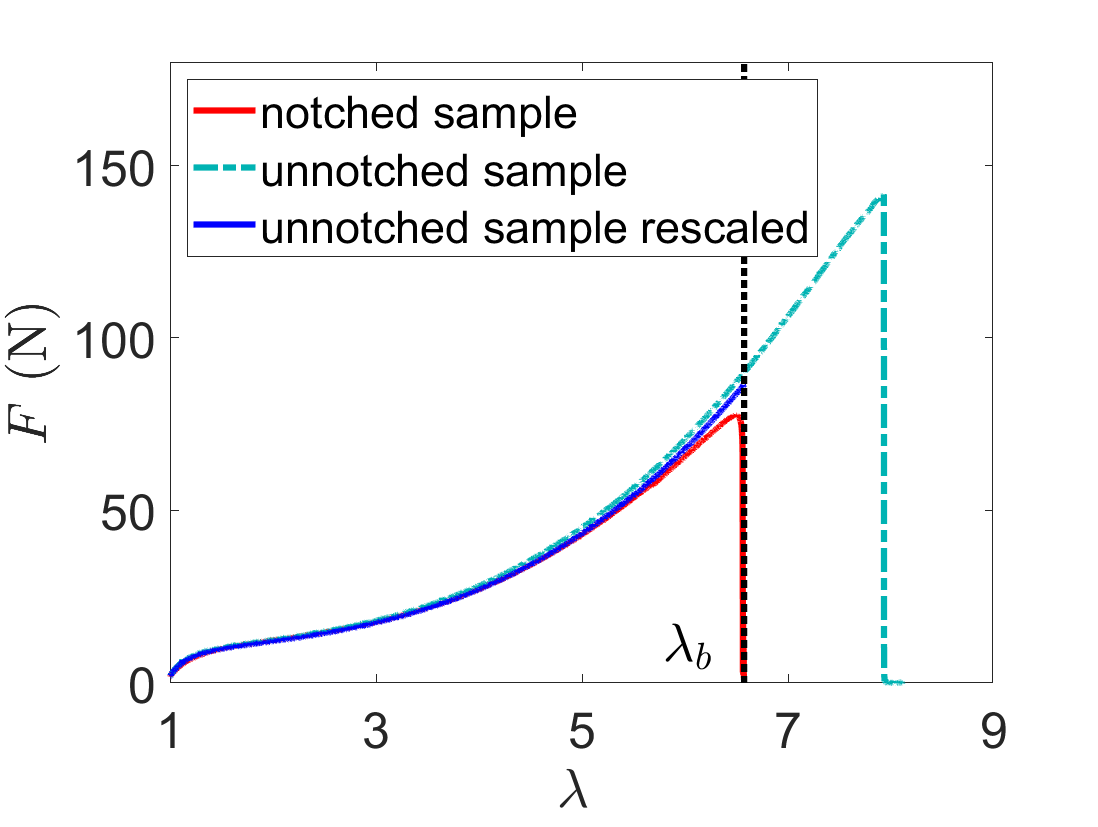}
\caption{Curve rescaling to match the initial elastic response of the notched sample to evaluate the fracture energy associated with crack deviation, example for $V = 5$~mm.s$^{-1}$ and $C = 17$\% (see text for details). }
\label{fig:FScl}
\end{figure}

To estimate the fracture energy specifically associated to sideways cracks, we determine the amount of mechanical energy released during the growth of the sideways crack. To that end, we compare the stored elastic energy in two sample configurations at a fixed stretch ratio $\lambda$. In the first configuration, a notched sample strained to  $\lambda$ stores an elastic energy $E_n(\lambda)$. This quantity cannot be directly measured if a crack has already initiated and begun to deviate at deformation $\lambda$. To estimate $E_n(\lambda)$, we use stress-strain data from tensile tests on unnotched samples and rescale the  curves to match the initial elastic response of the notched sample, prior any crack growth (Fig.~\ref{fig:FScl}). In the second configuration, a notched sample at the same deformation, but with a developed crack deviation, stores a smaller elastic energy, $E_d(\lambda)$. The difference in elastic energy $\Delta E(\lambda)= E_n(\lambda)- E_d(\lambda)$ represents the elastic energy dissipated due the growth of a crack deviation. At the rupture stretch ratio $\lambda_b$, and assuming the deviation extends over a distance $\ell_d$, we estimate the fracture energy associated with crack deviation as:
\begin{equation}
\Gamma_d = \frac{\Delta E(\lambda)}{\ell_d t}
\end{equation}
where $t$ is the sample thickness.

\subsection{Estimate of the elasto-adhesive length}

\begin{figure}[!htbp]
\includegraphics[width=0.42\linewidth]{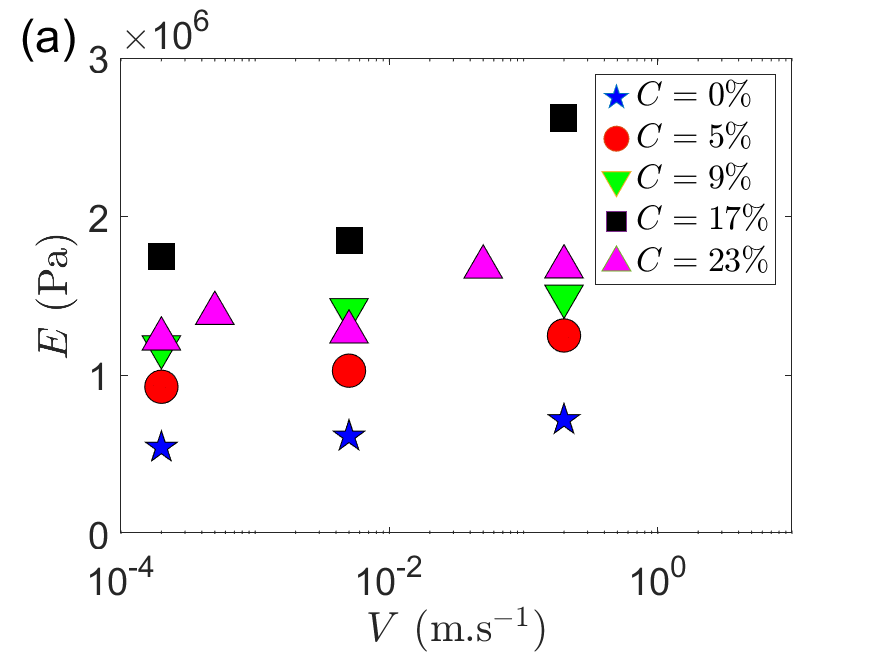}\includegraphics[width=0.42\linewidth]{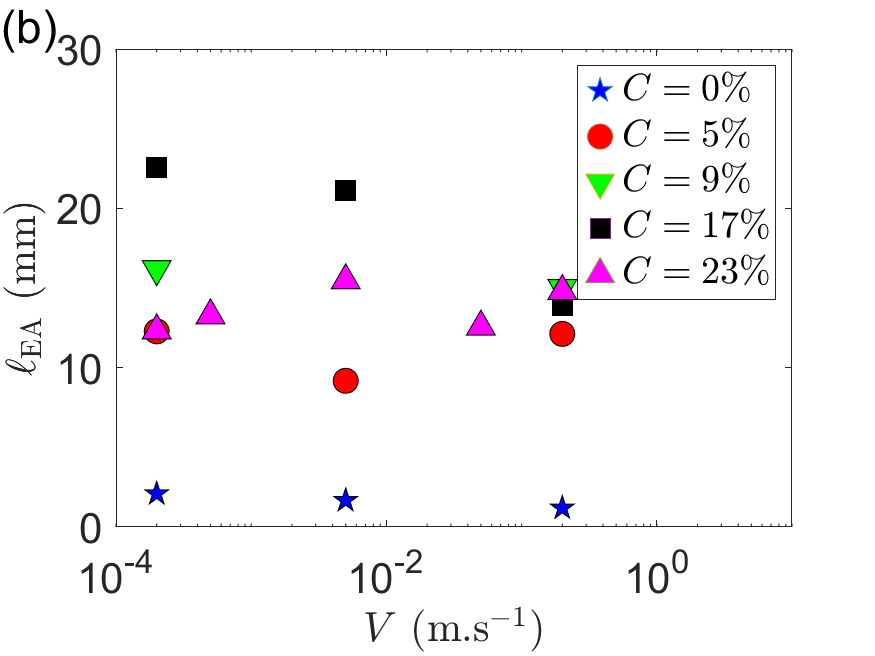}
\caption{(a) Elastic modulus $E$ as a function of loading rate $V$ for various filler fractions. (b) Elasto-adhesive lengths determined from the loading rate and filler dependent elastic modulus $E$ and fracture energy.}
\label{fig:ElEA}
\end{figure}

The elasto-adhesive length is defined as $\ell_{\mathrm{EA}}=\Gamma/E$, where $\Gamma$ is the filler-dependent fracture energy shown in Fig.~\ref{fig:MasterBreakCurve}(b) and $E$ is the small-strain Young's modulus. To determine $E$, we fit the low-strain portion of the stress-strain curves measured on unnotched samples using the hyperelastic Mooney-Rivlin model $\sigma=(2C_1+2C_2/\lambda)(\lambda-1/\lambda^2)$, and set $E=6(C_1+C_2)$. Because the stress-strain curves depend on the pulling velocity, the elastic modulus increases slightly and approximately logarithmically with pulling velocity, as commonly observed in polymers. For the filled formulations, the elasto-adhesive length is typically of the order of $10~\mathrm{mm}$ or larger, whereas it is approximately $1.5\pm 0.5~\mathrm{mm}$ for unfilled PDMS. In all cases, for a given filler fraction and pulling velocity, it exceeds the corresponding crack-deviation amplitude.

\end{document}